\documentstyle{mn}

\input epsf

\newif\ifAMStwofonts

\newcommand{\erg} {{\rm erg}\, {\rm cm}^{-2}\, {\rm s}^{-1}}
\newcommand{\keV} {{\rm keV}}
\newcommand{\Mpc} {{\rm Mpc}}

\newcommand{\dI}  {\left( {\Delta I\over I}\right)}
\newcommand{\R}   {{\it ROSAT}}
\newcommand{\A}   {{\it ASCA}}
\newcommand{\AX}  {{\it AXAF}}
\newcommand{\AB}  {{\it ABRIXAS}}
\newcommand{\XMM}   {{\it XMM}}
\newcommand{\HEAO}   {{\it HEAO-1 A2}}
\newcommand{\Ginga}  {{\it Ginga LAC}}
\newcommand{\RXTE}   {{\it RXTE}}
\newcommand{\cts}    {{\rm ct}\, {\rm s}^{-1}}

\ifoldfss
  \ifCUPmtlplainloaded \else
    \NewTextAlphabet{textbfit} {cmbxti10} {}
    \NewTextAlphabet{textbfss} {cmssbx10} {}
    \NewMathAlphabet{mathbfit} {cmbxti10} {} 
    \NewMathAlphabet{mathbfss} {cmssbx10} {} 
  \fi
  \ifAMStwofonts
    \ifCUPmtlplainloaded \else
      \NewSymbolFont{upmath} {eurm10}
      \NewSymbolFont{AMSa} {msam10}
      \NewMathSymbol{\upi}     {0}{upmath}{19}
      \NewMathSymbol{\umu}     {0}{upmath}{16}
      \NewMathSymbol{\upartial}{0}{upmath}{40}
      \NewMathSymbol{\leqslant}{3}{AMSa}{36}
      \NewMathSymbol{\geqslant}{3}{AMSa}{3E}

    \fi
  \fi
\fi 

\ifnfssone
  \newmathalphabet{\mathit}
  \addtoversion{normal}{\mathit}{cmr}{m}{it}
  \addtoversion{bold}{\mathit}{cmr}{bx}{it}
  \newmathalphabet{\mathbfit} 
  \addtoversion{normal}{\mathbfit}{cmr}{bx}{it}
  \addtoversion{bold}{\mathbfit}{cmr}{bx}{it}
  \newmathalphabet{\mathbfss} 
  \addtoversion{normal}{\mathbfss}{cmss}{bx}{n}
  \addtoversion{bold}{\mathbfss}{cmss}{bx}{n}
  \ifAMStwofonts
    \ifCUPmtlplainloaded \else
      \UseAMStwoboldmath
      \makeatletter
      \new@mathgroup\upmath@group
      \define@mathgroup\mv@normal\upmath@group{eur}{m}{n}
      \define@mathgroup\mv@bold\upmath@group{eur}{b}{n}
      \edef\UPM{\hexnumber\upmath@group}
      \new@mathgroup\amsa@group
      \define@mathgroup\mv@normal\amsa@group{msa}{m}{n}
      \define@mathgroup\mv@bold\amsa@group{msa}{m}{n}
      \edef\AMSa{\hexnumber\amsa@group}
      \makeatother
      \mathchardef\upi="0\UPM19
      \mathchardef\umu="0\UPM16
      \mathchardef\upartial="0\UPM40
      \mathchardef\leqslant="3\AMSa36
      \mathchardef\geqslant="3\AMSa3E
    \fi
  \fi
\fi 

\ifnfsstwo
  \DeclareMathAlphabet{\mathbfit}{OT1}{cmr}{bx}{it}
  \SetMathAlphabet\mathbfit{bold}{OT1}{cmr}{bx}{it}
  \DeclareMathAlphabet{\mathbfss}{OT1}{cmss}{bx}{n}
  \SetMathAlphabet\mathbfss{bold}{OT1}{cmss}{bx}{n}
  \ifAMStwofonts
    \ifCUPmtlplainloaded \else
      \DeclareSymbolFont{UPM}{U}{eur}{m}{n}
      \SetSymbolFont{UPM}{bold}{U}{eur}{b}{n}
      \DeclareSymbolFont{AMSa}{U}{msa}{m}{n}
      \DeclareMathSymbol{\upi}{0}{UPM}{"19}
      \DeclareMathSymbol{\umu}{0}{UPM}{"16}
      \DeclareMathSymbol{\upartial}{0}{UPM}{"40}
      \DeclareMathSymbol{\leqslant}{3}{AMSa}{"36}
      \DeclareMathSymbol{\geqslant}{3}{AMSa}{"3E}
    \fi
  \fi
\fi 

\ifCUPmtlplainloaded \else
  \ifAMStwofonts \else 
    \def\upi{\pi}
    \def\umu{\mu}
    \def\upartial{\partial}
  \fi
\fi

\title[Power Spectrum at high z from XRB]
{Measuring the Power Spectrum of Density Fluctuations at Intermediate
Redshift with X-ray Background Observations}
\author[X. Barcons et al]
       {X. Barcons$^{1,2}$, A.C. Fabian$^1$ and F.J. Carrera$^2$\\
	$^1$ Institute of Astronomy, Madingley Road, Cambridge CB3
0HA\\
        $^2$ Instituto de F\'{\i}sica de Cantabria (Consejo Superior
de Investigaciones Cient\'{\i}ficas - Universidad de Cantabria), 39005
Santander, Spain}

\pagerange{\pageref{firstpage}--\pageref{lastpage}}
\pubyear{1997}

\begin{document}

\maketitle

\label{firstpage}

\begin{abstract}
The precision of intensity measurements of the extragalactic X-ray
Background (XRB) on an angular scale of about a degree is dominated by
spatial fluctuations caused by source confusion noise.  X-ray source
counts at the flux level responsible for these fluctuations, $\sim
10^{-12}\erg$, will soon be accurately measured by new missions and it
will then be possible to detect the weaker fluctuations caused by the
clustering of the fainter, more distant sources which produce the bulk
of the XRB.  We show here that measurements of these excess
fluctuations at the level of $\dI\sim 2\times 10^{-3}$ are within
reach, improving by an order of magnitude on present upper
limits. Since it is likely that most (if not all) of the XRB will be
resolved into sources by AXAF, subsequent optical identification of
these sources will reveal the X-ray volume emissivity in the Universe
as a function of redshift.  With these ingredients, all-sky
observations of the XRB can be used to measure the power spectrum of
the density fluctuations in the Universe at comoving wavevectors
$k_c\sim 0.01-0.1 \Mpc^{-1}$ at redshifts where most of the XRB is
likely to originate ($z\sim 1-2$) with a sensitivity similar to, or
better than, the predictions from large-scale structure theories. A
relatively simple X-ray experiment, carried out by a large-area
proportional counter with a $0.5-2\deg^2$ collimated field-of-view
scanning the whole sky a few times, would be able to determine the
power spectrum of the density fluctuations near its expected peak in
wavevector with an accuracy better than 10 per cent.
\end{abstract}


\section{Introduction}

\subsection{The X-ray Background and the Large-Scale
Structure of the Universe}

Thirty-five years of study after its discovery by Giacconi et al
(1962), the X-ray Background (XRB) has proven to be a valuable tool in
the study of the Large-Scale Structure (LSS) of the Universe. A
significant fraction of the XRB (particularly at soft energies) is now
resolved into sources, the vast majority of which are
extragalactic. The deepest surveys carried out with \R\ (see, e.g.,
Hasinger 1996 for a recent review) resolved about 60 per cent of the
$0.5-2\, \keV$ XRB into mostly Active Galactic Nuclei (AGN) and other
X-ray luminous galaxies (particularly Narrow-Line X-ray Emitting
Galaxies-NLXGs). The X-ray volume emissivity of these objects rises
rapidly from redshift $z=0$ to $z=1-2$ above which it appears to
decline. It is at redshifts of 1-2 where the bulk of the resolved soft
XRB originates.  Preliminary work on \A\ observations (Georgantopoulos
et al 1997) hints that something similar is happening at harder
energies, although the fraction of the XRB resolved is lower and the
sources much more difficult to identify due to the limited spatial
resolution of \A.

Isotropy arguments at soft X-ray energies (see, e.g., Carrera, Fabian
\& Barcons 1997) show that a good deal of the unresolved fraction of
the soft XRB has to come from redshifts $z>1$.  It is then concluded
that the bulk of the XRB originates precisely at the epoch where the
largest present-day structures collapse and where different
cosmological models give the most different predictions.

Any cosmological model has to confront two basic boundary conditions:
(1) the Universe was very smooth at $z=1500$ where the Cosmic
Microwave Background (CMB) was produced; and (2) the Universe is very
lumpy and rich in structure today ($z=0$).  Density fluctuations in
the Universe grow linearly from the highly homogeneous CMB
recombination epoch until they become non-linear at low redshifts and
collapse to form the LSS we see today. The redshift at which a density
fluctuation becomes non-linear depends on the spatial scale of the
fluctuation and on the cosmological model, but for the most popular
models, this happens at a typical redshift $z\sim 1-5$ for scales of
$10-100\, h^{-1}\, \Mpc$ ($H_0=100\, h{\rm km}\, {\rm s}^{-1}\, {\rm
Mpc}^{-1}$, $q_0=0.5$ and $\Lambda=0$ are used throughout unless
explicitly stated).  Studies of the XRB as proposed here will add a
further constraint to cosmological models, since they will measure the
power spectrum of the fluctuations at comoving wavevectors $k_c\sim
0.01-0.1\, h\, \Mpc^{-1}$ with sufficient sensitivity to constrain
cosmological scenarios. In particular the parameters governing the
evolution of the power spectrum (e.g., $q_0$) could be constrained by
these observations.

Mapping the structure of the Universe at intermediate
redshifts with X-rays has the added bonus that the objects 
presumably occur in the highest density regions.  Indeed,
cosmic X-ray sources require strong gravity to be switched-on (either
very extended potential wells, as in galaxy clusters, or very deep
ones as in AGN), and therefore the X-ray sources existing at
intermediate redshifts could actually correspond to the objects that
first formed in the Universe at even earlier epochs.  A possible indication
of this is in the `bias factor' for X-ray sources for which there is
some evidence that it has a high value, at least for the nearby
luminous X-ray sources (see, e.g., Miyaji 1995 and Boughn, Crittenden
\& Turok 1997 who find values larger than 5). Although direct studies
of the clustering of X-ray selected AGN (Boyle \& Mo 1993, Carrera et
al 1997) show that these values might be too large for
lower-luminosity objects, bias factors of 2 or 3 might be likely for
X-ray sources. This issue will be solved with \AB\ when the dipole of
the distribution of sources is compared to the XRB dipole.

Parallel studies at other wavelengths are currently being used to
extract information about LSS at high redshift.  Deep galaxy surveys,
and especially the Canada-France Redshift Survey, are showing how the
galaxy-galaxy correlation function evolves up to redshifts close to 1
(Le F\`evre et al 1996).  However, scales of the order $\sim 100\,
h^{-1} \Mpc$ are beyond the scope of such work. High redshift
clustering is also being investigated by means of different classes of
QSO absorption systems (Fern\'andez-Soto et al 1996, Cristiani et al
1997).  None of these studies provides clear evidence on how
clustering in the Universe evolves.  Again, scales in excess of $10\,
h^{-1} \Mpc$ are not easily accessible by these studies. QSO
clustering appears to be a realistic way to map large scale structure
at high redshift, although the scales we are dealing with here are
also difficult to study with existing samples (see, e.g., Croom \&
Shanks 1996). However, analyses of the spatial structure of deep radio
surveys (at a redshift $z\sim 1$) do appear promising (Cress et al
1996, Loan, Wall \& Lahav, 1997).

Unless otherwise stated, the above cosmological parameters will be
used throughout and X-ray fluxes will be referred to the 2-10 keV bandpass.

\subsection{Deep versus wide-area surveys}

The next generation of X-ray instruments to be operating in the next
decade are mostly based on X-ray imaging with grazing-incidence X-ray
telescopes working up to energies $\sim 10\, \keV$. This is indeed
necessary to unveil the origin of the XRB, since it should not be
forgotten that most of the XRB energy density resides at $30\, \keV$,
whereas most of our knowledge on the source content is so far limited
to energies below $3\, \keV$ where only a few per cent of the total
energy budget is contained.  It is then essential to carry out imaging
surveys at as high an energy as possible and to identify the sources
dominating the source counts.

X-ray imaging telescopes are well-suited to medium and deep
surveys. Their limited effective collecting-area calls for long
integration times to reduce photon counting noise, resulting in very
deep images, eventually down to the confusion limit.  We discuss some
of these missions and their relation to the present work in the next
subsection. The source counts (the so-called log N - log S relation)
are then determined down to very faint fluxes in a small solid angle
and therefore with limited precision. If exposures are long enough so
that confusion noise dominates over photon counting noise, a
fluctuation analysis provides an extension of the log N -- log S curve
for almost another decade in flux downwards.

An alternative way of using an X-ray telescope with a suitable imaging
detector is by doing a shallow all-sky survey.  This was done by \R\
at soft X-ray energies (Voges 1993) and will be done by \AB\ up to $10
\, \keV$ (Friedrich et al 1996). Typical exposure times are then short
and photon counting noise is the limiting factor. Such an experiment
is able to produce a map with accurate positions of the brightest
sources over the whole sky. One can then extract much information
about the cosmographical distribution of the sources and the local LSS
of the Universe.

The method we propose here requires the whole sky to be surveyed (to
achieve the best statistics) to measure the power spectrum of the
density fluctuations (PS) , with negligible photon counting noise.  In
order to avoid the strongest structures associated with the galaxy,
only photon energies above 2 keV will be considered (see discussion in
section 4). A diffuse galactic component, amounting to $<10$ per cent,
associated to the galaxy has been detected (Warwick, Pye \& Fabian
1980, Iwan et al 1982), but it is expected to be smooth on scales of a
degree and therefore will not contribute to the fluctutaions.  We
therefore focus on non-imaging instruments in order to obtain a large
effective collecting area, such that in a typical exposure of about
100 s, confusion noise (i.e., the fluctuations in the sky brightness
caused by the presence of the sources) exceeds photon counting noise.
The field of view cannot be collimated to much less than $1\, \deg^2$
to maintain the required low photon counting noise level. The most
clear example of such an experiment was the \HEAO\ experiment, and, as
an instrument, the {\it Ginga} Large-Area Proportional Counter (LAC),
which, unfortunately for the purposes of this paper, did not carry out
an all-sky survey.

For a homogeneous distribution of sources, the sky brightness
fluctuations on scales of a few degrees will be dominated by
relatively bright ($\sim 10^{-12}\, \erg$) nearby sources (see section
2 for details).  However, clustering of sources, if strong enough, can
be visible out to more distant redshifts. The reason is that if
several or many distant faint X-ray sources cluster within the scale
of a field-of-view, they will produce a large enough signal in the
distribution of X-ray sky intensities. As it will be shown later, the
imprint of inhomogeneities in the distribution of X-ray sources on the
excess fluctuations is weighted by the redshift-dependence of their
X-ray volume emissivity. This means that fluctuations in excess of
confusion noise produced by the bright, just unresolved, sources will be
most detectable at the redshifts where the bulk of the XRB originates.

It should also be emphasized, and this has been proven by similar
analyses on existing data, that besides having a negligible photon
counting noise, what determines the sensitivity in the measurement of
excess fluctuations is the number of independent observations -- as it
will be shown later $\left({\Delta I\over I}\right)_{2\sigma}\propto
N_{obs}^{-1/2}$, $N_{obs}$ being the number of sky positions where the
XRB intensity has been measured.  All-sky coverage is then essential.

\subsection{X-ray missions relevant to the present work}

There are a number of previous, existing and planned missions which
have made or are expected to make decisive steps towards the goal that
is pursued here.  Among them, the \HEAO\ all-sky experiment has proven
to be the most useful for cosmological work (Boldt 1987), since it
provided all-sky coverage with small photon counting noise.  X-ray
\HEAO\ maps have been used to measure the XRB dipole (Shafer \& Fabian
1983, Shafer 1983, Lahav, Piran \& Treyer 1997), maybe also higher
order multipoles (Lahav, priv comm), the search for the Great
Attractor (Jahoda \& Mushotzky 1989), cosmography of voids (Mushotzky
\& Jahoda 1992) and superclusters (Persic et al 1990) and many other
cosmologically relevant issues.  Many things have been learned from
that experiment, and in particular that an absolute determination of
sky brightness requires a combination of two different fields-of-view.
It will be shown here that when other data become available, the
\HEAO\ observations will enable a decisive step forward to be made in
the measurement of the PS at intermediate redshift.

As mentioned before, the \Ginga , with a field-of view of $\sim
1^{\circ}\times 2^{\circ}$ and a larger effective area, could have
provided a very accurate measurement of the PS at high redshift if it
had carried out an all-sky survey.  Also, it was rather unfortunate
for the present purposes that the \Ginga\ had all collimators with the
same angular size and therefore the non-cosmic XRB had to be modelled
(as oposed to subtracted). Nevertheless, even with the data available, very
interesting constraints on LSS were found (Carrera et al 1991,
Carrera et al 1993).

In this paper we propose to measure or constrain the excess
fluctuations by improving the precision of the confusion noise
produced by relatively nearby bright sources. That means that the
source counts down to fluxes $< 10^{-12}\, \erg$ need to be obtained
with high accuracy.  This relation is only known today from
fluctuation analyses of \HEAO\ (Shafer 1983) and \Ginga\ data (Hayashida
1989, Butcher et al 1997). More recently 
\A\ surveys (Inoue et al 1996, Georgantopoulos et al 1997) have
measured the source counts at fluxes $\sim 10^{-14}\, \erg$. However, these
results are based on small solid angles and therefore the
normalisation of the source counts is uncertain.

A forthcoming mission that will define the log N
- log S relation at $10^{-12}\, \erg$ fluxes with the highest accuracy
is \AB . This happens because \AB\ will detect all sources in the sky
brighter than this flux. A discussion of  the accuracy in the
modelling of confusion noise enabled by \AB\ is presented in the next
Section.

The next most important mission for the proposed experiment is \XMM
. \XMM\ will not only define the log N - log S at fainter fluxes, but,
most importantly, will find the X-ray spectrum of the sources that
contribute both to the confusion noise and the excess
fluctuations down to very faint fluxes. \XMM\ will also be measuring
or constraining the power spectrum of the fluctuations at comoving
wavevectors $k_c\sim 0.1-1\, h\, \Mpc^{-1}$, thus complementing the
observations at larger scales.

Finally, \AX\ may resolve the whole XRB at $\sim 1$~keV. The superb
X-ray angular resolution combined with large optical telescopes will
provide a direct insight into the evolution of the X-ray volume
emissivity of the sources producing virtually all the XRB, which is
one of the key inputs to model excess fluctuations in terms of source
clustering.

\subsection{Organisation of the paper}

In the next section we parametrise the noise components and
sensitivities relevant to the measurement of excess fluctuations by a
collimated field-of-view proportional counter. Photon counting noise
is estimated in terms of cosmic and non-cosmic backgrounds, based on
\Ginga\ and \RXTE\ observations. Confusion noise is modelled according
to our best (rather inaccurate) knowledge of the source counts at
$10^{-12}\, \erg$. Particular attention is paid to the issue of the
uncertainties and biases that source variability might introduce in
that modelling. We show that $\left({\Delta I\over I}\right)_{2\sigma}\sim
2\times 10^{-3}$ might be reachable with the \HEAO\ maps when \AB\
source counts become available.

Section 3 presents the relation between excess fluctuations and PS.
Assuming that the 2-10 keV redshift evolution of the volume emissivity
is similar to the one associated to the resolved component of the soft
XRB (mostly contributed by QSOs and NLXGs), we demonstrate that
sensitive measurements of the PS at $z\sim 1-2$ on comoving
wavevectors $k_c\sim 0.01 - 0.1\, h\, \Mpc^{-1}$ can be achieved with
a few degree collimator, such as \HEAO.  This corresponds to a PS
signal below that expected from Cold Dark Matter models, which could
therefore be accurately tested when reliable information on the 2-10
keV X-ray volume emissivity (presumably from \AX\ optical
identification of deep fields) becomes available.

Section 4 presents a more detailed study on what could be achieved
with a new X-ray mission surveying the whole sky with a $\sim 0.5-2\,
{\rm deg}^2$ beam. Using expected future information on the X-ray
volume emissivity, such an experiment would be able to measure the PS
at intermediate redshift with great accuracy at comoving wavevectors
$k_c\sim 0.01-0.1 h \Mpc^{-1}$. Specifically, constraints on the
evolutionary parameters of the PS (and in particular $q_0$) at the
involved redshifts could be found.  Such a mission would provide other
scientific results which are also outlined in that section.

In Section 5 we summarize our results.

\section{Sensitivity of XRB observations to excess fluctuations}

For a uniform distribution of X-ray sources, the distribution of
spatial fluctuations in the XRB, when observed through a given beam,
can be predicted if the source counts and the noise components
(particularly photon counting noise) are known.  This is well
documented (Scheuer, 1957, 1974; Condon 1974, Shafer 1983) and has
been applied many times with success to derive X-ray source counts
from XRB fluctuations (Shafer 1983, Hamilton \& Helfand 1987, Barcons
\& Fabian 1990, Hasinger et al 1993, Barcons et al 1994, Butcher et al
1997).  Under the assumption of a uniform distribution of sources in
the sky and of confusion noise (i.e., the noise coming from the
presence or absence of sources) dominating over photon counting noise,
the distribution of XRB fluctuations (the so-called P(D) curve) is
only sensitive down to a flux where there is about one source per
beam, and its width (standard deviation) is dominated by the sources
of which there are a few (this number depends on the slope of the
source counts, see below) per beam. When photon noise is important,
the total intrinsic dispersion of the intensity distribution is then
\begin{equation}
\left( {\Delta I\over I}\right)_{intrinsic}={\left(
\sigma_c^2+\sigma_{ph}^2\right)^{{1\over 2}}\over I_{XRB}},
\end{equation}
where $\sigma_c$ is the confusion noise, $\sigma_{ph}$ is the
photon counting noise and $I_{XRB}$ is the XRB intensity.
As a
practical issue, we should emphasize that an overall fit to the P(D)
curve is mostly sensitive to its second moment, but it has the
advantage over the variance that it is not dominated by the brightest
unresolved sources (i.e., the tail of the P(D)), but by the sources
where there are a few per beam.

Now, when the sources are {\it not} uniformly distributed, but
are clustered in the sky, the situation is different.  The predicted
P(D) distribution needs then all of the $n$-point correlation
functions, for $n$ at least as large as the {\it total} average number
of sources per beam (Barcons 1992). There are indeed models for the
$n$-point correlation function of objects (e.g., a simple gaussian model
where all correlation functions beyond $n=2$ are zero, or a model
where there is a random distribution of clusters of sources, all of
them with the same average profile which determines the $n$-point
correlation functions), but observationally there is little knowledge
of them beyond $n=3$.

The contribution of clustered sources to the shape of the P(D) does
not affect only the sources brighter than the one-source-per-beam
level, but all sources equally.  Clusters of very faint sources
might produce significant dispersion in the XRB intensity.  As will
be shown later, the effect of clustering is weighted by the X-ray
volume emissivity as a function of redshift.

Fortunately, in most situations the P(D) shape is
dominated by the confusion noise of relatively bright sources, and the
effect of clustering is only a small correction. In those cases it has
been customary to parametrise the effect of clustering in terms of
`excess fluctuations', i.e., a small quantity $\dI$ to be quadratically
added to the intrinsic dispersion of the P(D) (eqn. 1). In what follows
we shall assume this approach.

The key point of this paper is based on the fact that the measurement
of the variance of an approximately gaussian distribution (we can use
that approximation for P(D) for this particular purpose) is
distributed as $\chi^2$.  Therefore the 2$\sigma$ uncertainty with which the
dispersion of the P(D) can be measured is $(2/N_{obs})^{1/2}$ times the
dispersion itself, where $N_{obs}$ is the number of independent
observations of the XRB intensity. That means that the 2$\sigma$
sensitivity at which excess fluctuations could be measured is
\begin{equation}
\left( {\Delta I\over I}\right)_{2\sigma} =\sqrt{2\over
N_{obs}}{(\sigma_c^2+\sigma_{ph}^2)^{{1\over 2}}\over I_{XRB}}
\end{equation}
If we have a beam with solid angle $\Omega \deg^2$, and the whole high
galactic latitude sky ($\mid b\mid > 20^{\circ}$) is used, then this
number is $\sim 0.01 \Omega^{1/2} \left({\Delta I\over
I}\right)_{intrinsic}$, which is why all-sky coverage is essential.

It should be emphasized that eqn. (2) shows the {\it maximum}
precision that can be reached and it requires knowledge of the
intrinsic dispersion (eqn. 1) to better than this figure. In what
follows we estimate these values under general grounds.

We assume a proportional counter with collimated field of view of
solid angle $\Omega \deg^2$, effective area $10^4\, A_4\, {\rm cm}^2$
with an energy bandpass from $\epsilon_1$ to $\epsilon_2$. For most
purposes we will use $\epsilon_1=2\, \keV$ and $\epsilon_2=10\,
\keV$. This instrument is assumed to scan the whole sky in 6 months,
so the typical exposure time will be of the order of $t=100\,
t_{100}\Omega^{1/2}\, {\rm s}$.

\subsection{Photon Counting Noise}

Proportional counters detect events from the cosmic XRB as well as
from particles crossing the detector.  These have different energy
spectra: the XRB has an energy  spectrum
$I_{XRB}\propto\epsilon^{-0.4}$, which should be folded through the
appropriate instrumental response (assumed as a constant effective
area here) to be
converted to counts, while
the particle background is often well approximated by a constant energy
dependence in the number of counts detected. If we write the total count-rate as the sum of these two terms
\begin{equation}
C_{Tot}=C_{XRB}+C_{DET}
\end{equation}
then
\begin{equation}
C_{XRB}=a_1\Omega A_4 \Psi(-0.4,\epsilon_1,\epsilon_2)
\end{equation}
and 
\begin{equation}
C_{DET}=a_2 A_4 \Psi(1,\epsilon_1,\epsilon_2)
\end{equation}
where we take into account bandpass corrections through the function
$\Psi$
\begin{equation}
\Psi(\beta,\epsilon_1,\epsilon_2)={\epsilon_2^\beta-\epsilon_1^\beta\over
10^\beta-2^\beta}
\end{equation}
The constants $a_1$ and $a_2$ are more or less universal for the same
type of orbit. We have estimated their values based on the \Ginga\ and
the \RXTE\ PCA.  For \Ginga\ ($A_4=0.4$, $\Omega=2$, $\epsilon_1=4$,
$\epsilon_2=12$), $C_{XRB}\sim 8 \cts$ and $C_{Tot}\sim 14\, \cts$
(see, e.g., Carrera et al 1993), while for \RXTE\ ($A_4=0.14$,
$\Omega=1$, $\epsilon_1=2$, $\epsilon_2=10$), $C_{XRB}\sim 2.5\, \cts$
and $C_{Tot}\sim 5\, \cts$ (K. Jahoda, private communication).  From
both instruments we find consistent values around $a_1\approx
a_2\approx 15-17\, \cts$.

The photon counting noise contribution to the dispersion of the P(D)
curve can then be estimated as
\[
\left( {\sigma_{ph}\over I_{XRB}} \right)=0.024\, \left( A_4\,
t_{100}\right)^{-{1\over 2}}\times  
\]
\begin{equation} 
\left( {1\over \Omega \Psi(-0.4,\epsilon_1,\epsilon_2)}
 +{\Psi(1,\epsilon_1,\epsilon_2)\over \Omega^2
\Psi^2(-0.4,\epsilon_1,\epsilon_2)}\right)^{{1\over2 }}
\end{equation}

The accuracy of this value depends crucially on how stable the
particle background is around the orbit and on whether it can be
subtracted rather than modelled (this requires collimators of various sizes).
An equatorial orbit minimizes variations due to the particle background.

\subsection{Confusion Noise}

In what follows we assume that the source counts dominating
the confusion noise follow a euclidean power law 
\begin{equation}
{dN\over dS}=K\,
\Omega\, (\gamma-1) {1\over S_0} \left( {S\over
S_0}\right)^{-\gamma}
\end{equation}
 where $\gamma=2.5$, $S_0$ is a reference flux
arbitrarily chosen to be $10^{-14}\, \erg$ (2-10 keV) and the
normalisation $K\, =300\, K_{300}$ sources $\deg^{-2}$ is the
number of sources per square degree brighter than $S_0$. Both the slope and the
normalisation of the source counts are consistent with the fluctuation
analysis carried out with the \Ginga\ (Butcher et al 1997) and also match
the deeper \A\ surveys (Inoue et al 1996, Georgantopoulos et al
1997), all of them referred to the 2-10 keV band. The spectrum of the
\Ginga\ fluctuations also showed that the sources responsible for the
confusion noise on $1\deg^2$ scales have a power-law spectrum with
energy spectral index $0.7$ and negligible absorption (Butcher et al 1997).

Confusion noise can be estimated following Condon (1974).  If beams
with an intensity more than $\Gamma\sigma$ above the mean are removed
since they will be identified as sources, the variance of the
remaining map can be estimated iteratively and the confusion noise
(defined as the flux equivalent to a 1$\sigma$ signal in the intensity
histogram) is 
\begin{equation}
\sigma_c=S_0\left( K\Omega{\gamma-1\over
3-\gamma}\Gamma^{3-\gamma}\right)^{{1\over\gamma-1}}
\end{equation}
which for a
typical value $\Gamma\approx 5$ and euclidean counts is 
\begin{equation}
\sigma_c\sim
1.6\times 10^{-12}\erg\, \left( K_{300}\Omega \right)^{{2\over 3}}.
\end{equation}
This confusion noise  produces a contribution
to the intrinsic dispersion of the P(D) curve 
\begin{equation}
{\sigma_c\over I_{XRB}}=0.13\, K_{300}\Omega^{-{1\over
3}}{\Psi(-0.7,\epsilon_1,\epsilon_2)\over
\Psi(-0.4,\epsilon_1,\epsilon_2)}
\end{equation}
which will have to be added in quadrature to eqn. (7) to find the
total intrinsic dispersion of the P(D) curve.

A very important point is to what degree of accuracy  this term can be
estimated, since it is likely to be the main limiting factor to
measurement of excess fluctuations.  It is indeed important to have
an all sky survey (as \AB\ will provide) in order to minimize the
statistical errors on this quantity. Over $8500\, K_{300}$ sources are
expected at high galactic latitudes down to a flux of $10^{-12}\,
\erg$ which would mean an accuracy of the order of 1 per cent in $K$ and
slightly better for $\sigma_c$.

However, there is the issue of the source variability. AGN, which are
supposed to be the dominant class of source at a flux $10^{-12}\,
\erg$ are known to vary substantially on all timescales.  Assuming
that sources vary independent of each other and with a similar
(flux-independent) amplitude, the measured source counts in a flux
limited sample are slightly different from the average source counts
(which are relevant to the P(D)). The main reason is that the
steepness of the source counts makes a fraction of the sources of
average flux below the detection threshold  contribute to the
source counts above the detection threshold.  Indeed some of the
sources of average flux above the detection threshold would
also be undetectable, but these are less numerous. The net
result is that the source counts in a flux limited sample are an
overestimate of the average source counts, and it is this last one
which matters for the confusion noise.

In order to quantify the variability effect on the confusion noise we
have simulated fluxes of sources, whose average values are drawn from
the source counts given by eqn. 8 down to $3\times
10^{-13}\erg$. The number of simulated sources is set to cover the
whole high galactic latitude sky in order to mimic the \AB\ survey.
The fluxes are then allowed to vary randomly within a factor of several. 
That produces a different list of source fluxes which is then cut
at $10^{-12}\, \erg$. The source counts are then fitted via maximum
likelihood to a single power law (as in eqn.  8), 
resulting in an estimate of the
confusion noise given in eqn. 9.  

\begin{figure}
\vbox to 0cm{\vfil}
\epsfverbosetrue
\epsfysize= 200pt
\epsffile{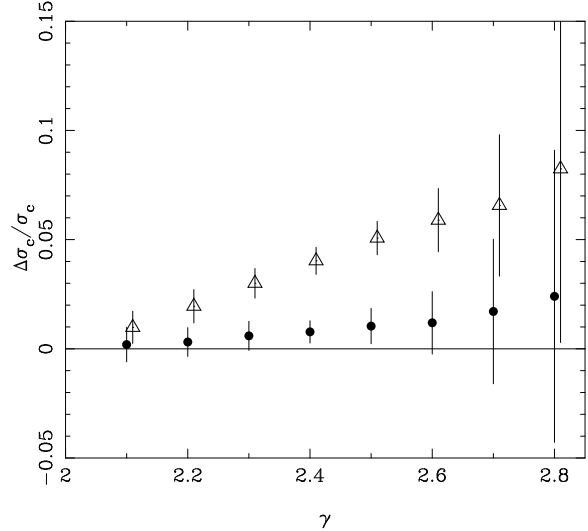}
\caption{Relative shift in the estimates of confusion noise from an
 all-sky survey as a function of the slope of the differential source
 counts. Filled dots correspond to a factor of 2 variability and
 triangles to a factor of 5 variability (these have been slightly
 shifted to higher values of $\gamma$ to avoid confusion between error
 bars). The error bars represent the
 standard dispersion due to statistics and variability.  }
\end{figure}

Fig. 1 displays the relative variation between fitted and expected
confusion noise as a function of input slope $\gamma$ for $K_{300}=1$
when all sources vary within a factor of 2 and a factor of 5.  The
amount of overestimation of the confusion noise is $\sim 1$~ per cent
for euclidean counts if sources vary within a factor of 2 and, as
expected, it grows with $\gamma$.  Since not all of the sources vary
(there will be some contribution in the source counts from clusters of
galaxies, galaxies, etc.) the error is likely to be smaller.  For
euclidean counts the error is therefore small enough to reach the
maximum accuracy in the excess fluctuations, assuming source
variability within a factor of 2.  The situation is clearly worse if
sources vary within a factor of 5, where for euclidean counts the
variation in the confusion noise will be of the order of 5 per
cent. There are indications (R.S. Warwick, private communication) that
source variability could be large on average (close to a factor of 5)
from the comparison of {\it Rosat} All-Sky Survey data with later
pointed observations.  However, since variability produces a
systematic effect on the confusion noise, different observations
carried out by \AB\ (e.g., observations taken 6 months apart) of the
same sources could be used to quantify
this effect and to correct the source counts.

\subsection{Sensitivity to excess fluctuations}

Eqn. 1, when combined with eqn. 11, shows that the minimum intrinsic
dispersion in the P(D) curve for a $\sim 1$~deg set of observations is
around 10 per cent.  The accuracy with which any excess variance could
be measured on top of this intrinsic value depends on both the
precision with which this dominating term can be modelled. If the
intrinsic dispersion can be modelled to better than 1 per cent (which
is close to the maximum statistical accuracy in the absence of source
variability), then excess variances as low as $\sim 10^{-3}$ could be
detected.  

More precisely, and taking into account the estimated values for the
confusion and photon counting noise from the previous subsections and
assuming that source variability is not going to dominate the
precision with which confusion noise can be estimated, the excess
fluctuations that could be eventually measured by an all-sky
observation with a collimated field-of-view proportional counter are
shown in Fig. 2.  This assumes a 2-10 keV bandpass.

\begin{figure}
\vbox to 0cm{\vfil}
\epsfverbosetrue
\epsfysize= 200pt
\epsffile{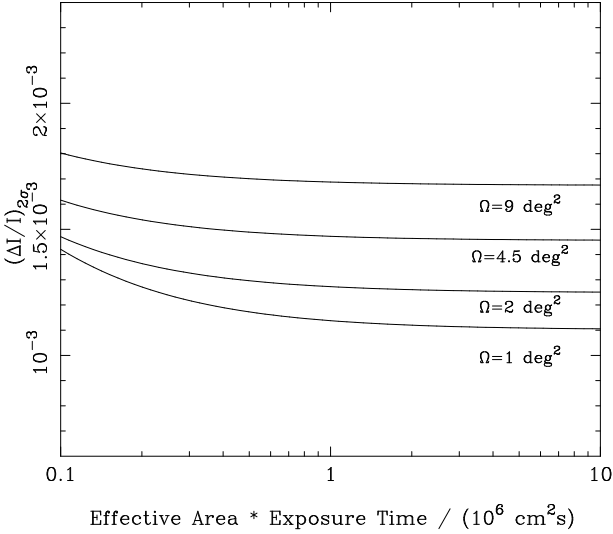}
  \caption{Minimum detectable excess variance for all-sky
  observations and different beam sizes}
\end{figure}

Clearly, when the product $A_4t_{100}$ is larger than a few (the
precise value depending on the beam $\Omega$), the
intrinsic dispersion of P(D) is dominated by confusion noise as
opposed to the photon counting noise which  dominates at
smaller values. When P(D) is confusion
dominated, excess variances as small as $2\times 10^{-3}$ could be
measured even with relatively large beams.  For $\Omega=1$, values
close to $1\times 10^{-3}$ would be within reach.

\section{Measuring the Power Spectrum of the Fluctuations at High
Redshift}

The excess variance can be related to the power spectrum of the
density fluctuations in the
following way (see Barcons \& Fabian 1988, Carrera, Fabian \& Barcons
1997 for details):
\[
\left( {\Delta I\over I}\right)^2 = {1\over
4\sqrt{2\pi}}{c\over H_0 I_{XRB}^2}\int dz\, F(z;q_0)
\]
\begin{equation}\int d^2q\, \hat G^2(q) \left(
{2\over\pi}\right)^{{3\over 2}}{\cal P}(z;q/d_A(z))
\end{equation}
where the XRB intensity $I_{XRB}$ is 
\begin{equation}
I_{XRB}={\Omega_{eff}\over
4\pi}{c\over H_0}\int dz\, (1+z)^{-5}(1+2q_0z)^{-{1\over 2}}j(z).
\end{equation}
$\Omega_{eff}=3.046\times 10^{-4}\Omega$, $j(z)$ is the X-ray volume 
emissivity at redshift $z$ (with the appropriate
K-correction), $d_A$ is the angular distance 
\begin{equation}
d_A(z)={c\over
H_0}{\left[ zq_0 +(q_0-1)(-1+\sqrt{1+2q_0z})\right]\over q_0^2 (1+z)^2},
\end{equation}
$\hat G(q)$ is the 2D Fourier transform of the beam function and
${\cal P}(z;k)$ is the power spectrum of the fluctuations ($k\equiv
(1+z)k_c$ is the {\it physical} wavevector) , which is related to the
source 2-point correlation function $\xi(z;r)$ by
\begin{equation}
\left( {2\over\pi}\right)^{{3\over 2}} {\cal P}(z;k)={1\over
(2\pi)^{{3\over 2}}}\int d^3r\, e^{-i\vec{k}.\vec{r}}\xi(z;r),
\end{equation}
and finally
\begin{equation}
F(z;q_0)=(1+z)^{-8}(1+2q_0z)^{-{1\over 2}}j^2(z)/d_A^2(z).
\end{equation}
The above
equations provide the basic link
between excess fluctuations, evolution in the X-ray volume emissivity
(the actual normalization is irrelevant since it cancels out)
and the power spectrum. Although the PS enters in the expression of
the excess fluctuations in a convoluted way, there are a couple of
simplifications that provide an 
almost one-to-one relation between power spectrum and
excess fluctuations. 

In what follows a circular beam with gaussian profile (dispersion
angle $\theta_0$) will be assumed.

\begin{figure}
\vbox to 0cm{\vfil}
\epsfverbosetrue
\epsfysize= 200pt
\epsffile{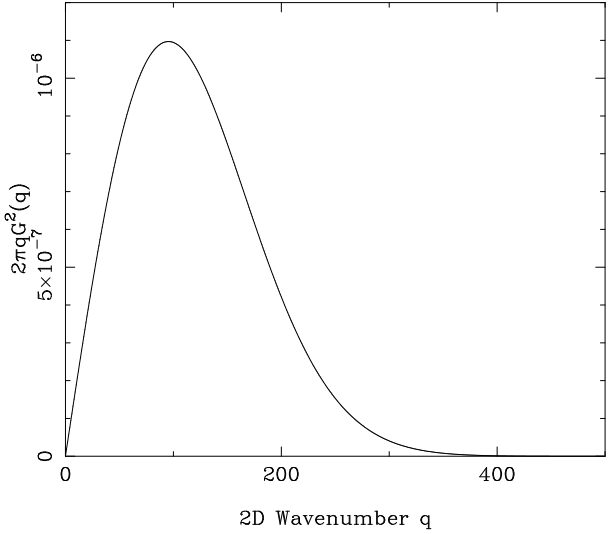}

  \caption{Effective beam filtering function for a $1^{\circ}$ FWHM
  circular beam with gaussian profile as a function of the 2D
  wavevector.}
\end{figure}

The first simplification to realise is that only wavevectors $k\sim
\Omega_{eff}^{-{1\over 2}}\, d_A(z)^{-1}$ are relevant. The effective
filtering function $2\pi q\hat G^2(q)$ is shown in Fig.~3 for a
$1\deg$ FWHM beam (i.e., $\theta_0=1/2.354\, \deg$). It can be seen
that only values of the 2D wavevector $q$ around the maximum
$q_{max}=2^{-{1\over 2}}\theta_0^{-1}\sim 100\Omega^{-{1\over 2}}
\deg^{-1}$ will contribute to the excess fluctuations. That implies
that at every redshift $z$, only the 3D wavevectors $k\sim
q_{max}/d_A(z)$ will contribute substantially to the excess
fluctuations. For beamsizes of the order of a degree, and significant
redshifts $z\sim 1$, the 3D wavevectors to which it is maximally
sensitive are $k\sim 0.01-0.1\, h\, \Mpc^{-1}$, which is
near the expected maximum of the PS. Then, it is safe to assume that
the PS does not vary much over the range of relevant wavevectors,
which results in the following expression for the excess fluctuations
\begin{equation}
\left( {\Delta I\over I}\right)^2=\int dz  {\cal W}(z){\cal
P}( z;(2^{{1\over 2}}\theta_0d_A(z))^{-1}),
\end{equation}
where
\begin{equation}
{\cal W}(z)={2H_0\over c\Omega_{eff}}{(1+z)^{-8}(1+2q_0z)^{-{1\over
2}}j^2(z)/d_A^2(z)\over\left[\int
dz'(1+z')^{-2}(1+2q_0z')^{-{1\over 2}}j(z')\right]^2}.
\end{equation}

The next simplification is hinted at by our (limited) knowledge of the
redshift evolution of the X-ray volume emissivity $j(z)$. 
It is hoped that after \AX\ and \XMM\ are launched, and deep
surveys have been carried out, optical identification work (especially for
the \AX\ sources whose positions will be determined with very good
accuracy) will be able to reveal the volume emissivity $j(z)$ as a
function of redshift.  So far, at soft energies it appears that AGNs
have a steeply rising volume emissivity ($j(z)\propto K(z)(1+z)^{3+p}$,
$K(z)$ being the K-correction and $p\sim 3$) up to $z_c=1.7$, but beyond
that redshift everything is consistent with no evolution (Boyle et al
1994, Page et al 1996).  The NLXGs that appear to be more numerous at
faint fluxes, might also show a similar behaviour (Griffiths et al
1996).  To our present knowledge then, $j(z)$ is a steeply rising
function up to some redshift $z_c$ and then it flattens out to a
constant comoving volume emissivity.

In practice that means that the influence of the power spectrum
(i.e. inhomogeneities in the distribution of sources) on the excess
fluctuations is heavily weighted towards $z_c$ for wavevectors $k<0.1 h
\Mpc^{-1}$. In order to illustrate this fact, we define the function $W(k,z)$
\begin{equation}
\left( {\Delta I\over I}\right)^2=\int
dk\, \int dz \,  W(k,z){\cal P}(z;k)
\end{equation}
Fig. 4 shows $W(k,z)$ for different values of $k$ assuming $p=3$, an
energy spectral index $\alpha=1$ for the sources to compute the
K-correction (the actual energy spectral index is likely to be smaller
and therefore the function $W(k,z)$ will be much more peaked towards
$z_c$) and $z_c=1.7$. What is seen there is that the wavevectors that
dominate are around $k\sim 0.1 h\, \Mpc^{-1}$ and that within these,
it is the redshift at which the volume emissivity peaks that is most
heavily weighted. There is of course some contribution from smaller
wavevectors at lower redshifts, but since the peak in the power
spectrum is expected to be found around these shorter wavevectors, not
much contamination from large-scale local structure is to be
expected. In fact, this large-scale power could be removed by
`flat-fielding' the all-sky maps with a large-scale smoothed version
of the same maps, leaving scales of  $k\sim 0.1\, h\, \Mpc^{-1}$
unaffected. This would have the additional advantage of removing any
residual galactic large-scale structure. It is then concluded that we
could be measuring the PS at a redshift beyond the deepest available galaxy
surveys.
\begin{figure}
\vbox to 0cm{\vfil}
\epsfverbosetrue
\epsfysize= 200pt
\epsffile{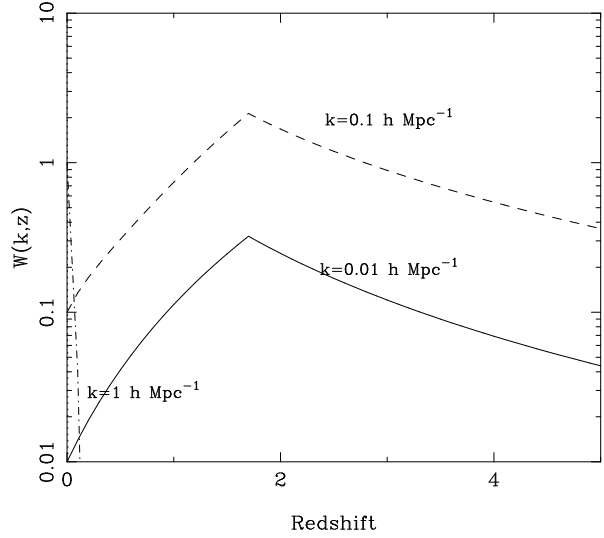}
  \caption{The weight $W(k,z)$ defined in eqn. 19 as a
  function of redshift for representative values of $k$}
\end{figure}

To estimate the sensitivity in the measurement of the PS with X-ray
observations, we assume that all of the XRB comes from a redshift bin
$\Delta z=2$ around $z_c=1.7$ and approximate the integrals in
redshift as the central value of the integrand times $\Delta z$. 
We then find 
\begin{equation}
\left( {\Delta I\over I}\right)^2={2\over\Delta V}{\cal P}(z_c;k_0),
\end{equation}
where $k_0=2^{-{1\over 2}}\theta_0^{-1}d_A(z_c)^{-1}$ and $\Delta V$
is the volume sampled by a beam
\begin{equation}
\Delta V=\Omega_{eff}\, d_A(z_c)^2\,
cH_0^{-1}(1+z)^{-2}(1+2q_0z)^{-{1\over 2}}\Delta z.
\end{equation}

\begin{figure*}
\vbox to 0cm{\vfil}
\epsfverbosetrue
\epsfysize= 300pt
\epsffile{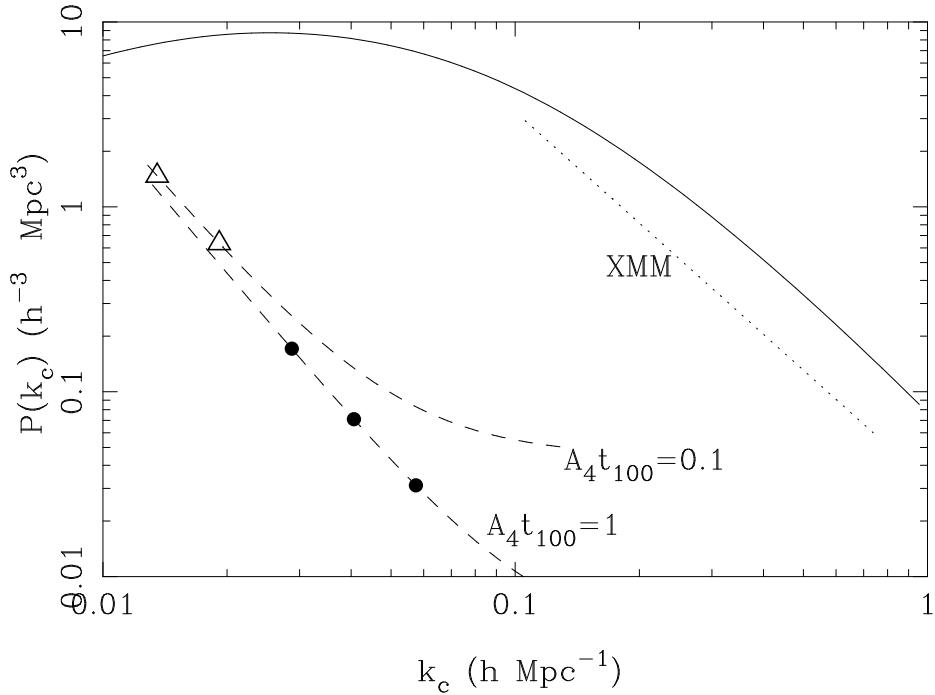} 
 \caption{Power spectrum of density fluctuations as a function of
  comoving wavevector all computed at a redshift $z=1.7$. The solid line
  represents the prediction from local observations evolved with $q_0=0.5$.
  The dashed curves are the $2\sigma$ sensitivities from all-sky
surveys carried out with various beam sizes and for the quoted values
of the parameter $A_4t_{100}$.  The triangles show the maximum
sensitivity that could be reached with \HEAO\ data (for the
$3^{\circ}\times 3^{\circ}$ and $1.5^{\circ}\times 3^{\circ}$
collimators from left to right) and the filled points show the
sensitivity of the instrument whose concept is presented in Section
4.1, with beam sizes of 2, 1 and 0.5 deg (left to right).  The dotted
line shows the sensitivity of XMM observations for 2 years as
explained in the text.}
\end{figure*}

Fig.  5 shows the maximum sensitivity, in terms of the power spectrum
as a function of comoving wavevector, for different beam sizes,
according to the estimates from Section 2.  A factor of 2 reduction in
the number of independent measurements has been included in order to
account for the fact that neighbouring observations will not be
independent. For the \HEAO\ points (triangles) $A_4t_{100}=0.1$.  The
dashed lines represent the expected maximum sensitivities for all-sky
observations of the XRB with different beam sizes (ranging from $10\,
{\rm deg}^2$ to $0.1\, {\rm deg}^2$ and various values of
$A_4t_{100}$. The filled dots represent beams of 2, 1 and 0.5 deg$^2$
for $A_4t_{100}=1$, as in the instrument whose concept is presented in
the next section. From this figure it is clearly seen that the best
sensitivity near the expected peak of the PS is achieved by 0.5-1
deg$^2$ beams and that to avoid the results being dominated by photon
counting noise (i.e., the flattening of the dashed curves towards high
wavevectors) at those angles, a value of $A_4t_{100}$ close to 1 is needed.

In the same figure we also show the expected sensitivity reached at
larger wavevectors with \XMM\ observations over 2 years.  That has
been computed assuming a log N -log S as observed at soft X-ray
energies with $\gamma=2$ but with a normalisation two times larger, as
it seems to apply to the 2-10~keV passband (Georgantopoulos et al
1997). We have assumed about 500 useful observations with an average
exposure time of 20 ks during that period.

We also show for comparison a Cold Dark Matter power spectrum (see
Peacock 1997 and in particular Fig. 6 of that paper) linearly evolved
to redshift 1.7 (with $q_0=0.5$). A shape parameter $\Gamma^*=0.25$
has been assumed, claimed by Peacock \& Dodds (1994) to fit the shape
of the local LSS very well and the normalisation has been chosen
accordingly.  The conclusion is that when the confusion noise from
sources brighter than $\sim 10^{-12}\, \erg$ can be accurately
modelled, the all-sky \HEAO\ observations might be sensitive enough to
measure the PS at intermediate redshift (that is in the absence of
other systematics). With a smaller beam (close to $0.5-1\, {\rm
deg}^2$) and larger area the accuracy in the measurement could be
of a few per cent.

\section{Measuring the XRB fluctuations on 1 degree scale}

\subsection{A mission dedicated to the XRB}

Throughout this paper we have illustrated what results would be
obtained by using a beamsize close to $1^{\circ}$. In what follows we
want to show how the simplest possible instrument (i.e. a collimated
field-of-view proportional counter) could provide a very significant
cosmological result.  

Indeed other instruments might provide in principle similarly valuable
information with comparable sensitivity in terms of excess variance.
For values of the product $A_4t_{100}>0.1$, the measurement of the
fluctuations could be sensitive enough to provide cosmologically
relevant results.  We believe that systematics are going do dominate
the ultimate sensitivity of such measurements.  If unpredictable and
slow gain variations in the instrument are the dominant source of the
systematics, then the larger the effective area the better, especially
when the measurement can be repeated a few times. A detailed study
would be required for each experiment where systematics need to be
understood and kept to a minimum.  We devote here special attention to
proportional counters which are well understood and and have proven
stable over extended periods.

Although the smallest detectable excess fluctuations $\left({\Delta I\over
I}\right)_{2\sigma}$ for a $1\, \deg^2$ expriments would be within a factor of 2 of
what can be done with \HEAO\ (see Fig. 2), the sensitivity in the
measurement of the power spectrum would be increased by a much larger
factor, since a much smaller volume $\Delta V$ would be sampled by a
single beam (see eqn. 20).  To further illustrate this point we show
in Fig.~6 what such an instrument would be able to do in terms of
measuring the power spectrum, as a function of the parameter
$A_4t_{100}$ and beam $\Omega$. Fig. 6 also shows the expected values
for a 1-5 keV bandpass which would collect more counts, but due to the
fact that the sources that dominate the confusion noise have a steeper
spectrum than the XRB, the P(D) curves would be noisier and the
experiment less sensitive.

\begin{figure}
\vbox to 0cm{\vfil}
\epsfverbosetrue
\epsfysize= 200pt
\epsffile{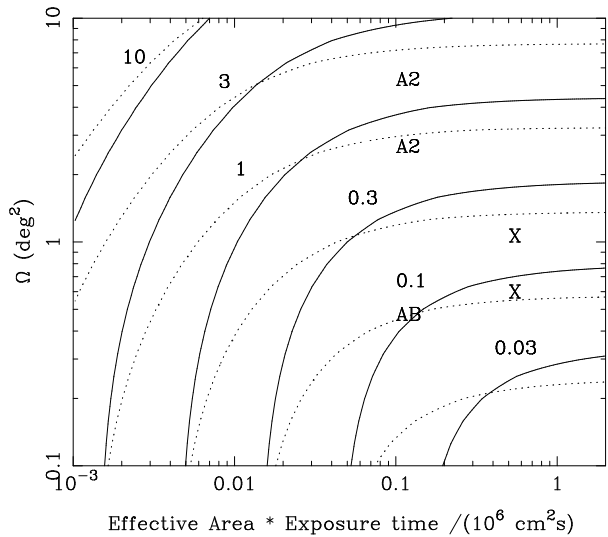} 

 \caption{Contour levels of constant power spectrum (in $h^{-3}\,
 \Mpc^{3}$ units) achievable at redshift $z_c=1.7$. The solid lines
 represent the 2-10 keV bandpass and the dotted lines the 1-5\, keV
 bandpass.  The 2 \HEAO\ collimator sizes (4.5 and 9 $\deg^2$) have
 been labelled ``A2'', while the instrument proposed here has been
 labelled ``X'' (collimator sizes 1 and 2 $\deg^2$). The {\it ABRIXAS}
 point, assuming an area of 25${\rm cm}^2$ and a total integration
 time of 4000 s is shown as `AB'.}
\end{figure}

We propose an instrument of effective area $A_4=1$, with two collimator
sizes (1 and 2 $\deg^2$) so the contribution from cosmic and detector
backgrounds could be well separated. It might be actually very
interesting that the collimators are elongated (e.g.,
$0.5^{\circ}\times 2^{\circ}$ and $1^{\circ}\times 2^{\circ}$) in
which case there will be some information on the power spectrum down
to smaller scales.
With such a large effective area,
the X-ray brightness of the sky at any point could be determined with
an accuracy better than 2 per cent. The power spectrum at $k_c\sim
0.01-0.1\, h\, \Mpc^{-1}$ could then be measured, in principle, down
to $0.1 h^{-3} \Mpc^{3}$ or better (2$\sigma$) which would guarantee not only a
detection (if the PS at intermediate redshift is not largely
overestimated by the above parametrisation) but also an accurate
measurement.  In what follows we try to highlight the need for such an
instrument and also list a few complementary studies that would
benefit considerably from such observations.

First of all, Figs. 5 and 6 show the minimum detectable power spectra,
assuming that the intrinsic width of the P(D) can be determined to its
absolute statistical precission.  Indeed, it might be that source
variability is found to have a larger effect or that photon counting
noise cannot be modelled to 1 per cent accuracy. If one or both of
these effects cause a 5 per cent uncertainty in the intrinsic P(D)
noise, then with the \HEAO\ experiment (degrading its sensitivity in
the power spectrum by almost an order of magnitude) the cosmic signal
in the PS would be missed. A 5 per cent accuracy in the intrinsic P(D)
width with the proposed experiment would nevertheless allow a power
spectrum as small as $\sim 2 h^{-3} \, \Mpc^{3}$ (still below
the prediction) to be detected at 2$\sigma$. Indeed, the fact
that such an instrument would be working close to the peak of the
power spectrum helps.

\begin{figure}
\vbox to 0cm{\vfil}
\epsfverbosetrue
\epsfysize= 200pt
\epsffile{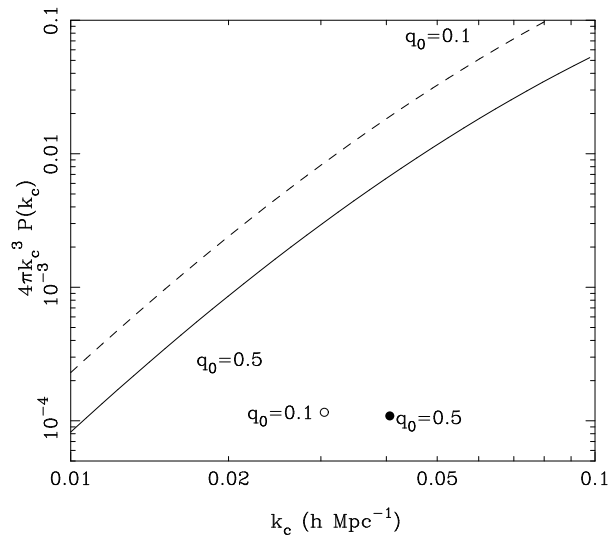} 

 \caption{Dimensionless power spectrum estimated at different values
 of $q_0$.  The solid curve corresponds to a linearly evolved
 $q_0=0.5$ local PS (see text). The dashed line shows the same thing
 for $q_0=0.1$. The filled and empty dots show the expected $2\sigma$
 sensitivities from the observations suggested in Section 4.1 and for 
 $q_0=0.5$ and $q_0=0.1$ respectively.}
\end{figure}

To further illustrate the capabilities of this observation, we
explored the possibility that a sensitive measurement of the PS at
intermediate redshift could constrain the evolutionary parameters of
the PS.  We have taken the $z=0$ Cold Dark Matter spectrum with
constant $\Gamma^*$ (see Section 3 and Peacock 1997) and evolved it
linearly (which is appropriate to the scales under consideration) to
$z=1.7$ with different values of $q_0$. The dimensionless power
spectrum ($4\pi k_c^3{\cal P}(k_c)$, Fig. 7) does not depend on the
Hubble constant $h$, but does depend on $q_0$.  The expected $2\sigma$
accuracy for the dimensionless PS is between one and two orders of
magnitude smaller than the predictions. The difference between the
expected dimensionless power spectrum for $q_0=0.1$ and $q_0=0.5$ is
only a factor of 4 at that redshift and therefore $q_0$ could be
measured in principle.

Such a mission makes no severe technical demands requiring only a
continuous scan of the whole sky over 2 years with the most stable
proportional counters and 1-axis stabilisation.  In order to keep the
contribution to the intrinsic dispersion of the P(D) curve well
determined, it is necessary to have the detector background as stable
as possible.  In this respect, the mission concept we propose here
would be better suited if it is launched into an equatorial orbit (as
for BeppoSAX).

\subsection{Other ways to measure Large-Scale Structure}

There are many other scientific goals that could be achieved while
such an instrument is performing its main task.  Within the subject of
the LSS, the multipoles of the XRB should be mentioned first. In fact
at least the dipole needs to be well measured and compared with the
dipole in the distribution of the X-ray sources to estimate the bias
parameter.  The earliest attempts to measure LSS with the use of the
XRB looked also for 12-hr and 24-hr effects which could be attributed
to the presence of a large lump of X-ray emitting matter at different
distances (Warwick, Pye \& Fabian 1980). \HEAO\ provided the
possibility of a measurement of the dipole of the XRB, whose direction
is in general agreement (within large errors) with the CMB dipole
(Shafer 1983, Shafer \& Fabian 1983).  The amplitude of the XRB dipole
is expected to be larger than the CMB one primarily because of
aberration ($\left( {\Delta I\over I}\right)_{\rm
Dip}=(3+\alpha){v\over c}$, $\alpha$ being the energy spectral index
which is $\alpha \sim 0.4$ for the XRB and $\alpha=-2$ for the CMB),
but also because the mass overdensity which is gravitationally pulling
the Local Group is also expected to emit X-rays above the average
(Warwick, Pye \& Fabian 1980, Miyaji 1995, Miyaji \& Boldt 1990 and
Boldt 1992).

More recently Lahav, Piran \& Treyer (1997) have proposed a formalism
to measure multipoles in the XRB and to relate them to the power
spectrum of the fluctuations, as is done with galaxy surveys.  Of
course the main problem in measuring multipoles is not a sensitivity
one, but confusion noise.  The amplitude of the dipole is  $<1$~per cent
and must be measured in maps where confusion noise is
larger. Since all the multipoles are in fact variances of the sky maps
weighted with appropriate spherical harmonics, the contribution of
confusion noise is dominated by the brightest sources that have not
been removed. Than makes the measurement of multipoles particularly
difficult (Lahav, private communication). In addition there is the
unknown contribution of the galaxy, even at high galactic latitudes
which could contribute to the low-order (large-scale) multipoles.

There is however an advantage in the experiment proposed here with
respect to the only previous one that carried out an all-sky survey
(\HEAO ), when combined with \AB .  The `shot noise' variance for all
multipoles is (see Lahav, Piran \& Treyer 1997) 
\begin{equation}
\langle \mid
a_{lm}\mid^2\rangle_{sn}=S_0^2\, K\, \Omega {\gamma-1\over 3-\gamma}
\left( {S_{max}\over S_0}\right)^{3-\gamma}
\end{equation}
where $S_{max}$ is the flux above which all sources have been removed.
If \AB\ can find the positions of all sources brighter than
$10^{-12}\erg$, of surface density is $0.3\, \deg^{-2}$, then one out
of every 3 beams will have to be excluded from the multipole analysis.
This still provides enough data, with virtually no impact on the
amplitude of the multipole signal, but with the `shot noise' reduced
by a factor of 5, when compared with the \HEAO\ maps where the
Piccinotti et al (1982) sources have been removed.

The proposed mission will also provide a good measurement of the
Autocorrelation Function. Since the instrument would scan the sky
along great circles, measurements of the sky brightness at separations
less than the collimator field-of-view will be taken.  These can be
used to measure the autocorrelation function of the XRB on scales
smaller than the beam size. However, since the beams will strongly
overlap in adjacent measurements, the signal in the Autocorrelation
Function will be dominated by the brightest sources present (see
Carrera et al 1993 and Mart\'\i n-Mirones et al 1991). At the very
least, this will be useful to confirm the source counts at high fluxes
(as was done, for example, in the Ginga High Galactic Latitude Survey
by Kondo 1991).  Beyond the angular scale where the beams overlap, the
autocorrelation function is expected to measure true source clustering
and extension of the cosmic sources. If clusters of galaxies can be
reliably removed from the all-sky maps (again using \AB ), source
clustering on scales larger than the beam size could also be measured.

Other possibilities for such an instrument include the search for
positive or negative signals around known structures (Great Attractor,
Superclusters, Voids, etc.), studies of the cross-correlation function
between XRB intensity and galaxy and cluster catalogues (Jahoda et al
1991, 1992; Lahav et al 1993a, Miyaji et al 1994, Barcons et al 1995,
Carrera et al 1995, So{\l}tan et al 1996) as well as with X-ray maps
at softer energies, cross-correlations of the XRB with Cosmic
Microwave Background maps (Boughn \& Jahoda 1993; Boughn, Crittenden
\& Turok, 1997, Kneissl et al 1997) possible detection of excess
skewness in the fluctuations (similar to what is done in
counts-in-cells, Lahav et al 1993b) and many types of studies of the
distribution of X-ray sources and diffuse emission from the Galaxy.

\section{Conclusions}

We have shown that all-sky degree-scale observations of the XRB
carried out with enough sensitivity (effective area times exposure
time in excess of $10^6\, {\rm cm}^2{\rm s}$) could reveal excess
fluctuations due to the clustering of distant sources down to levels
$\dI \sim 1-2\times 10^{-3}$. These excess fluctuations are likely to
arise from high redshift sources and then the power spectrum of the
fluctuations at those early epochs could be measured with sufficient
sensitivity to detect the cosmic signal.

The generic requirement for such a goal is that the intrinsic width of
the P(D) curve (caused by confusion noise and photon counting noise)
is estimated to better than 1 per cent. Source variability (if
within a factor of 2) already introduces a bias of 1 per cent for
euclidean source counts if an all-sky survey of sources brighter than
$10^{-12}\, \erg$ is used (as will be carried out by \AB ). If
variability is as large as a factor of 5, then the contribution of
confusion noise to the intrinsic P(D) dispersion will suffer from a 5
per cent error.  Besides this, photon counting noise needs to be
modelled to better than 1 per cent to achieve the maximum sensitivity
in the excess fluctuations.

When the X-ray source counts down to that flux have been measured, the
P(D) noise in the \HEAO\ maps could be accurately modelled and the
power spectrum of the density fluctuations at intermediate redshift
might be measurable if the modelling of P(D) can be done accurately
enough.

A new X-ray mission
consisting of a large effective area proportional counter with 1 and 2
$\deg^2$ collimators, which would scan the whole sky several times
would provide a much more secure approach.
Such an experiment would have a sensitivity 10 times better in terms
of the power spectrum than the \HEAO\ experiment and would be able to
detect a signal in the power spectrum over 10-100 times smaller than the
predictions of the popular models. We believe that such an
experiment constitutes the best chance to measure the power spectrum
of the density fluctuations in the Universe at a redshift $z\sim 1-2$
where the different cosmological scenarios give the most distinct
predictions.

\section*{Acknowledgments}

We want to thank Elihu Boldt, G\"unther Hasinger, Keith
Jahoda, Ofer Lahav and Bob Warwick for their comments about the
concept presented in Section 4.1, which resulted in substantial
improvements which propagated throughout the paper.
Partial financial support for XB and FJC was provided by the DGES under
project PB95-0122. Financial help for XB's sabbatical at Cambridge was
provided by  DGES grant PR95-490. ACF thanks the Royal Society for support.

\bsp

\label{lastpage}


\begin{thebibliography}{99}



\bibitem{b01} Barcons, X. \& Fabian, A.C. 1988, MNRAS, 230, 189
\bibitem{b02} Barcons, X. \& Fabian, A.C., 1990, MNRAS, 243, 366
\bibitem{b03} Barcons, X., 1992, ApJ, 396, 460
\bibitem{b04} Barcons, X. et al 1994, MNRAS, 268, 833
\bibitem{b05} Barcons, X. et al 1995, ApJ, 455, 480
\bibitem{b06} Boldt, E., 1987, PhysRep, 145, 215
\bibitem{b07} Boldt, E., 1992, In: The X-ray Background, eds. X. Barcons \&
A.C. Fabian, Cambridge University Press,  p. 115
\bibitem{b08} Boughn, S.P., Jahoda, K., 1993, ApJ, 412, L1
\bibitem{b09} Boughn, S.P., Crittenden, R.G., Turok, N.G., 1997, New
Astronomy, submitted
\bibitem{b10} Boyle, B.J. \& Mo, H.J., 1993, MNRAS, 260, 925
\bibitem{b11} Boyle, B.J., Shanks, T., Georgantopoulos, I., Stewart,
G.C., Griffiths, R.E.,  1994, MNRAS, 271, 639
\bibitem{b12} Butcher, J.A., Stewart, G.C., Warwick, R.S., Fabian,
A.C., Carrera, F.J., Barcons, X., Hayashida, K., Inoue, H., Kii, T.,
1997, MNRAS, in the press
\bibitem{b13} Carrera, F.J., Barcons, X., Butcher, J.A., Fabian, A.C.,
Stewart, G.C., Warwick, R.S., Hayashida, K., Kii, T., 1991, MNRAS, 249, 698
\bibitem{b14} Carrera, F.J., Barcons, X., Butcher, J.A., Fabian, A.C.,
Stewart, G.C., Toffolatti, L., Warwick, R.S., Hayashida, K., Inoue,
H., Kii, T., 1993, MNRAS, 260, 376
\bibitem{b15} Carrera, F.J., Barcons, X., Butcher, J.A., Fabian, A.C.,
Lahav, O., Stewart, G.C., Warwick. R.S., 1995, MNRAS, 275, 22
\bibitem{b16} Carrera, F.J.,  Fabian, A.C. \& Barcons, X., 1997,
MNRAS, 285, 820
\bibitem{b17} Carrera, F.J., et al , 1997, in preparation
\bibitem{b18} Condon, J.J., 1974, ApJ, 188, 279
\bibitem{b19} Cress, C.M., Helfand, D.J., Becker, R.H., Gregg, M.D., White, R.L., 1996, ApJ, 473, 7
\bibitem{b20} Cristiani, S., D'Odorico, S., D'Odorico, V., Fontana,
A., Giallonago, E., Savaglio, S., 1997, MNRAS, 285, 209
\bibitem{b21} Croom, S.M. \& Shanks, T., 1996, MNRAS, 281, 893
\bibitem{b22} Fern\'andez-Soto, A. et al 1996, ApJL, 460, 85
\bibitem{b23} Friedrich, P., Hasinger, G., Richter, G., Kritze, K.,
Tr\"umper, J., Br\"auninger, H., Predehl, P., Staubert, R.,
Kendizorra, E., 1996, In: R\"ontgenstrahlung from the Universe,
eds. Zimmermann, H.U., Tr\"umper, J., Yorke, H.; MPE Report 263, p. 681
\bibitem{b24} Le F\`evre O., Hudon, D., Lilly, S.J., Crampton, D.,
Hammer, F., Tresse, L., 1996, ApJ, 461, 534
\bibitem{b25} Georgantopoulos, I., Stewart, G.C., Blair, A.J., Shanks,
T., Griffiths, R.E., Boyle, B.J., Almaini, O., Roche, N., 1997, MNRAS, in the press
\bibitem{b26} Giacconi, R., Gursky, H., Rossi, B., Paolini, F., 1962,
Phys. Rev. Lett., 9, 
\bibitem{b27} Griffiths, R.E., Della Ceca, R., Georgantopoulos,
I. Boyle, B.J., Stewart, G.C., Shanks, T., Fruscione, A., 1996, MNRAS, 281, 71
\bibitem{b28} Hamilton, T.T. \& Helfand, D.J., 1987, ApJ, 318, 93
\bibitem{b29} Hasinger, G., Burg, R., Giacconi, R., Hartner, G.,
Schmidt, M., Tr\"umper, J., Zamorani, G.,  1993, A\&A, 175, 1
\bibitem{b30} Hasinger, G., 1996, A\&AS, 120, C607 
\bibitem{b31} Hayashida, K., 1989, PhD Thesis, Kyoto University
\bibitem{b32} Inoue, H., Kii, T., Ogasaka, Y., Takahashi, T., Ueda,
Y., 1996, 
In: R\"ontgenstrahlung from the Universe,
eds. Zimmermann, H.U., Tr\"umper, J., Yorke, H.; MPE Report 263,
p. 323
\bibitem{b33} Iwan, D., Shafer, R.A., Marshall, F.E., Boldt, E.A.,
Mushotzky, R.F., Stottlemeyer, A., 1982, ApJ, 260, 111
\bibitem{b34} Jahoda, K. \& Mushotzky, R.F.,  1989, ApJ, 346, 638
\bibitem{b35} Jahoda, K. et al 1991, ApJ, 378, L37; Erratum: 1992, ApJ, 399, L107 
\bibitem{b36} Kneissl, R., Egger, R., Hasinger, G., So{\l}tan, A.M.,
Tr\"umper, J., 1997, A\&A, 320, 685
\bibitem{b37} Kondo, H., 1991, Ph D Thesis, University of Tokyo
\bibitem{b38} Lahav, O., Fabian, A.C., Barcons, X., Boldt, E.,
Butcher, J.A., Carrera, F.J., Jahoda, K., Miyaji, T., Stewart,
G.C. Warwick. R.S., 1993a, Nat, 364, 693
\bibitem{b39} Lahav, O., Itoh, M., Inagaki, S. \& Suto, Y., 1993b, ApJ, 402, 387
\bibitem{b40} Lahav, O., Piran, T. \& Treyer, M.A., 1997, MNRAS, 284, 499
\bibitem{b41} Loan, A., Wall, J.V., Lahav, O., 1997, MNRAS, 286, 994
\bibitem{b42} Mart\'\i n-Mirones, J.M., De Zotti, G., Franceschini,
A., Boldt, E.A., Marshall, F.E., Danese, L., Persic, M., 1991, ApJ, 379, 507
\bibitem{b43} Miyaji, T. \& Boldt, E., 1990, ApJL, 353, 3
\bibitem{b44} Miyaji, T. et al  1994, ApJ, 434, 424
\bibitem{b45} Miyaji, T. 1995, PhD Thesis, Univ of Maryland 
\bibitem{b46} Mushotzky, R.F. \& Jahoda, K.,  1992, In: The X-ray
Background, eds. 
X. Barcons \& A.C. Fabian, Cambridge University Press, p. 80
\bibitem{b47} Page, M.J., et al 1996, MNRAS, 281, 579
\bibitem{b48} Persic, M., Jahoda, K., Rephaeli, Y., Boldt, E.,
Marshall, F.E., Mushotzky, R.F., Rawley, G., 1990, ApJ, 364, 1
\bibitem{b49} Piccinotti G., Mushotzky, R.F., Boldt, E.A., Holt, S.S.,
Marshall, F.E., Serlemitsos, P.J., Shafer, R.A., 1992, ApJ, 253, 485 
\bibitem{b50} Scheuer, P.A.G., 1957, Proc Camb Phil Soc, 53, 764
\bibitem{b51} Scheuer, P.A.G., 1974, MNRAS, 166, 329
\bibitem{b52} Shafer, R.A., 1983, PhD Thesis, Univ of Maryland
\bibitem{b53} Shafer, R.A. \& Fabian, A.C., 1983, IAU Symp 104, p. 333
\bibitem{b54} So{\l}tan, A.M., Hasinger, G., Egger, R., Snowden, S.,
Tr\"umper, J., 1996, A\&A, 305, 17 
\bibitem{b55} Voges, W., 1993, Adv Space Res, 131,391
\bibitem{b56} Warwick, R.S.,  Pye, J.  \& Fabian, A.C.,  1980, MNRAS, 190, 243


\end{thebibliography}
\end{document}